\documentclass[aps,prb,showpacs,amsmath,amssymb,
unsortedaddress]
{revtex4}
\usepackage{graphicx}

\newcommand{\q}{\mathbf{q}}
\newcommand{\p}{\mathbf{p}}
\renewcommand{\k}{\mathbf{k}}

\renewcommand{\r}{\mathbf{r}}

\newcommand{\beq}{\begin{equation}}
\newcommand{\eeq}{\end{equation}}
\newcommand{\Q}{\mathbf{Q}}
\newcommand{\beqa}{\begin{eqnarray}}
\newcommand{\eeqa}{\end{eqnarray}}

\begin{document}

\title{Strong-coupling theory of magnetic-exciton-mediated
superconductivity in UPd$_2$Al$_3$}

\author{P. McHale}
\affiliation{Max-Planck-Institut f\"{u}r
Physik komplexer Systeme, N\"{o}thnitzer Str. 38, 01187 Dresden, Germany}
\author{P. Thalmeier}
\affiliation{Max-Planck-Institut f\"{u}r
Chemische Physik fester Stoffe, N\"{o}thnitzer Str. 40, 01187 Dresden, Germany}
\author{P. Fulde}
\affiliation{Max-Planck-Institut f\"{u}r
Physik komplexer Systeme, N\"{o}thnitzer Str. 38, 01187 Dresden, Germany}
\date{\today}

\begin{abstract}

There is compelling evidence from 
inelastic-neutron-scattering and tunneling experiments 
that the heavy-fermion superconductor UPd$_2$Al$_3$ 
can be understood as a dual system consisting
of magnetic excitons, arising from crystal-field-split
U$^{4+}$ levels, coupled to delocalised f-electrons. 
We have computed the superconducting
transition temperature and the mass renormalisation arising from a
dual model with maximal spin anisotropy using a strong-coupling approach.
We find an instability to two possible opposite-spin-pairing states
with even- or odd-parity gap functions.
Each has a line node perpendicular to the c-direction, in agreement with
NMR relaxation-rate, specific-heat and thermal-conductivity measurements.
In addition, both
have total spin component $S_z$=0, compatible with the observation of
a pronounced Knight shift and $H_{c2}$ Pauli limiting. 
For parameter values appropriate to UPd$_2$Al$_3$,
we determine the dependence of the superconducting transition temperature
$T_c$ on a phenomenological coupling constant $g$ and we
investigate the associated mass enhancement and its anisotropy.
 
\end{abstract}


\maketitle

\section{Introduction}

Heavy-fermion superconductors are complicated materials which possess
a fascinating and rich variety of physical properties 
and which have stimulated the creation of 
an equally diverse range of theories in an effort to understand
them. These materials are dominated by strong electronic correlations giving
rise to magnetic and superconducting instabilities. 
Accordingly, most theories ignore phononic degrees of freedom. 

One of the most extensively studied models of heavy-fermion superconductivity
is a phenomenological model based on the exchange of spin fluctuations between
heavy quasiparticles. The latter originate in periodic resonant
scattering of conduction electrons close to the Fermi level as
described within the Kondo lattice model of Ce compounds
\cite{HewsonBook,Fulde88}.
Pairing via spin fluctuations can be considered as 
a one-component model to the extent that the spin fluctuations originate in the
system of heavy quasiparticles and any interaction between
conduction electrons and more localised electrons, or conduction
electrons and phonons, is neglected. In spite of its simplicity this
model has proved 
to be very useful in our attempts to understand heavy-fermion compounds on 
the border of magnetic long-range order, such as 
CePd$_2$Si$_2$\cite{Mathur98}, CeIn$_3$
\cite{Mathur98}
and CeRhIn$_5$\cite{Curro03}. Its success probably
derives from the 
fact that the spin-fluctuation-induced effective interaction tends 
to dominate all other channels of interaction when a
material is tuned close to the border of magnetism by doping or,
as in
the Ce compounds mentioned above,  
by applying pressure. 
The spin-fluctuation mechanism in the Kondo lattice may also be appropriate
for the Ce-based heavy-fermion superconductors at ambient pressure, 
CeCu$_2$Si$_2$ and CeCoIn$_5$\cite{Petrovic01,Curro03}.

It has become increasingly evident that the Kondo-lattice model
is, however, not adequate in the case of heavy-quasiparticle formation
in some Uranium heavy-fermion 
compounds \cite{Zwicknagl02,Zwicknagl03}, where 5f electrons
are partly localised and partly itinerant. The former occupy 
crystalline-electric-field (CEF) split 5f$^2$ states. 
The latter, more itinerant, 5f
electrons have a strongly enhanced effective mass due to 
a coupling to virtual
excitations between CEF states. This notion of heavy-quasiparticle formation is
especially appropriate for UPd$_2$Al$_3$ and it is the starting point
of our theoretical model
for heavy-fermion 
superconductivity in this compound. Specifically, we consider a
novel effective pairing mechanism which is based on the virtual
exchange of collective CEF excitations known as `magnetic
excitons'. They are propagating bosonic modes in contrast to the overdamped
modes which give rise to  pairing 
in the itinerant spin-fluctuation model.

The interplay of conduction electrons and CEF excitations is a well
studied subject. Initially, research was focused in two 
areas\cite{FuldePeschel72}: transport anomalies and
superconducting pair breaking or enhancement due to
conduction-electron scattering from dilute CEF-split impurities;
and periodic lattices of CEF ions interacting with one another
via conduction-electron polarisation. In the latter case attention was
focused on collective effects within the CEF system due to the
RKKY interactions mediated by itinerant electrons.
Later, White and Fulde\cite{WhiteFulde81} showed that the inverse effect,
viz. conduction-electron mass enhancement via virtual emission and
absorption of magnetic excitons, is also important. They demonstrated
that this mechanism explains the enhanced effective electron masses in 
praseodymium metal. 

It is then natural to ask whether the exchange of
magnetic excitons between quasiparticles can also mediate
superconductivity. Until recently, 
no theoretical work had been carried out to answer this
question, mainly for two reasons. 
Firstly, when the effects on superconductivity 
of paramagnetic impurities with CEF-split 
energy levels were studied in detail\cite{Fulde70},
it was found that for rare-earth ions the pair-breaking transition matrix
elements usually dominate the pair-enhancing matrix elements. 
Secondly, no good
example of a superconducting compound with this dual nature, i.e.,
magnetic excitons
arising from localised f electrons coupled to delocalised f electrons, was 
known. 
Recently, however, evidence has been accumulating that
the heavy-fermion
compound UPd$_2$Al$_3$ ($T_N = 14.3$K, $T_c = 1.8$K, see
Ref.~\onlinecite{Geibel91}) is the first example where such a 
mechanism is responsible for superconductivity.
Since this mechanism is pair breaking in the 
s-wave channel (as was already known from the impurity models mentioned above),
the superconducting gap function must change sign as a function
of $\k$ (i.e., the gap must have a node) as is the case with
spin-fluctuation-mediated superconductivity. 

An early indication that UPd$_2$Al$_3$ is a localised-delocalised f system
came from magnetic-susceptibility measurements.
Grauel et
al\cite{Grauel92} measured the dc magnetic susceptibility 
in UPd$_2$Al$_3$ and argued that the anisotropy which they observed arose 
from a tetravalent configuration of the
Uranium ions, i.e. U$^{4+}$ ($5f^2$). 
Since then Knight-shift\cite{Feyerherm94} 
and optical\cite{Dressel02}
measurements have supported the notion that UPd$_2$Al$_3$ contains both
localised and delocalised f electrons. 

However, 
the most direct evidence in favour of the dual-system hypothesis comes from
inelastic neutron scattering (INS) and tunnelling experiments. 
A dispersive crystal-field excitation (magnetic exciton) was
observed by Mason and Aeppli\cite{Mason97} in INS	
experiments within the AF phase. Later higher-resolution  
experiments\cite{Sato97,Bernhoeft98,Bernhoeft00} revealed a resonance feature
which appears in the INS spectrum upon entering the superconducting state.  
This result demonstrates that a strong interaction exists between the
localised and delocalised components of the f-electron system. Pioneering 
tunnelling experiments\cite{Jourdan99} performed on UPd$_2$Al$_3$-AlO$_x$-Pb
tunnelling junctions allowed experimenters to view, for the first time, the
tunnelling density of states (DOS) of a heavy-fermion
superconductor. Strong-coupling features appear in the DOS around 1meV close 
to the superconducting gap energy reinforcing the view that the exchanged
bosons are the magnetic excitons. 
Taken together, INS and tunnelling experiments 
lead Sato and co-workers\cite{Sato01} 
to the conclusion that superconductivity arises in
this material from an 
effective interaction between itinerant electrons mediated by magnetic
excitons. Using a model two-component Hamiltonian they were able to
explain qualitative features of the INS scattering spectrum and the
superconducting tunnelling spectrum. 

Subsequently, the origin of the magnetic excitons and their global
dispersion, as measured in Ref.~\onlinecite{Mason97}, was investigated in more
detail\cite{Thalmeier02}. Using an xy-type interaction, where 
only the $\sigma_\pm$ components of the conduction-electron spin
couples to the magnetic excitons, the effective non-retarded pair
potential was derived. The gap equations were solved within a
weak-coupling
approach demonstrating that the 
highest $T_c$ is obtained by an odd-parity 
state. In this model, however, the structure of the pairing amplitudes in
spin space is complicated, 
making it unsuitable for going beyond the non-retarded approximation.

In this paper, therefore, we investigate an alternative model with a
simplified interaction
between localised and itinerant 5f electrons. This interaction 
is of the Ising
type, i.e., only the $\sigma_z$ component of the conduction-electron spin can scatter magnetic excitons. In this case, the gap
equations naturally divide into those for equal- and opposite-spin
pairing, in contrast to the usual `singlet' and `triplet' classification
that arises in spin-rotation-symmetric models. 
In the present work we treat this simplified dual model for UPd$_2$Al$_3$  
in a more sophisticated strong-coupling approach using a mapping to an
electron-boson  Hamiltonian and solving the \'{E}liashberg equations for the
frequency- and momentum-dependent self-energy and gap functions. 
The level scheme which we use
here, and which gives rise to the Ising-type interaction, is quite realistic:
recent band-structure calculations based on this level scheme
account well for
experimental de Haas-van Alphen frequencies in
UPd$_2$Al$_3$\cite{Zwicknagl03}. 
We demonstrate that our 
model can yield a superconducting transition temperature and a mass
renormalisation which are consistent with experiment 
for reasonable values of the coupling constant. 

\section{Model}

Band structure calculations based on the supposition that two of the
three Uranium 5f-electrons  are localised reproduce the
observed de Haas-van Alphen frequencies in UPd$_2$Al$_3$ very 
well\cite{Zwicknagl03}. These calculations 
suggest a level scheme for the localised U 5f states which we adopt here.
According to the jj-coupling scheme the U$^{4+}(5f^2)$ ions have total
angular momentum $J=4$.
The twofold degeneracy of the ionic ground state is
lifted by a crystalline electric field. We consider only the excitation
between the non-degenerate ground state 
$|\Gamma_4\rangle = \frac{1}{2} \left( |J_z=3\rangle-|J_z=-3\rangle\right)$ 
and the first excited state 
$|\Gamma_3\rangle = \frac{1}{2} \left( |J_z=3\rangle+|J_z=-3\rangle\right)$. 
The CEF energy splitting is of the order $\Delta\simeq$ 6 meV,
as obtained from fitting the magnetic-exciton dispersion obtained from
INS \cite{Mason97} measurements to theoretical results \cite{Thalmeier02}. 

Although such a model can give rise to induced magnetism via mixing of
the states $|\Gamma_3\rangle$ and $|\Gamma_4\rangle$, we take 
the view here that the underlying antiferromagnetic (AF) 
order is not an important
consideration in the description of heavy-fermion superconductivity in
UPd$_2$Al$_3$. AF order mainly leads to a folding down of the conduction
bands into the AF Brillouin zone. Also, the magnetic-exciton dispersion in
the ordered phase is not appreciably different from that in the paramagnetic 
phase, due to the fact that
T$_N\ll\Delta$ \cite{Thalmeier02}. Furthermore, 
in mean-field theory, the AF order
parameter will only slightly modify the superconducting T$_c$ due to a
reconstruction of the conduction-electron states close to the AF Bragg
planes.

Accordingly we consider a three-dimensional lattice of localised
5f$^2$ CEF states and itinerant f-electrons. In the subspace
\{$|\Gamma_3\rangle,|\Gamma_4\rangle$\} we may write the CEF part of the 
Hamiltonian as
\beq
H_{CEF}=\Delta\sum_i S_{iz},
\eeq

\noindent
where ${\bf S}$ denotes a pseudospin (S=$\frac{1}{2}$). In this 
representation we interpret the CEF ground state as having
S$_z$= -$\frac{1}{2}$  and energy $-\Delta/2$ and the excited CEF state as
having  S$_z$= $\frac{1}{2}$ and energy $\Delta/2$. 
In the
pseudospin representation,
the only non-zero component of the physical total angular momentum ${\bf J}$
is the $J_z$-component: 
\beqa
J_z & = & \gamma
\left(
\begin{array}{cc}
0 & \frac{1}{2}\\
\frac{1}{2} & 0 
\end{array}
\right)
 = \gamma S_x ,
\eeqa

\noindent where $\gamma = 6$. 
Then the full two-component Hamiltonian may be written as
\beq
H = \sum_{\k,\sigma} \epsilon_\k c_{\k,\sigma}^\dagger c_{\k,\sigma} 
+ \Delta \sum_i S_{iz} - J \sum_{i\delta} S^i_x S^{i+\delta}_x
- I \sum_i \sigma_{iz} S_{ix} .
\label{HfirstVers}
\eeq

\noindent 
The third term in H 
is a nearest-neighbour 
superexchange interaction between localised 5f states. The
last term is the exchange interaction between localised 5f CEF states and the
delocalised 5f conduction electrons; 
the conduction-electron spin operator is
\beq
\mbox{\boldmath $\sigma$}_{i} = \psi_\alpha^\dagger(\r_i) 
\mbox{\boldmath $\sigma$}_{\alpha\beta} \psi_\beta(\r_i) . 
\eeq

\noindent
After carrying out a Holstein-Primakoff transformation (valid at low
temperatures $T \ll \Delta$) and an additional Bogoliubov transformation
involving the resulting Holstein-Primakoff 
bosons (see the Appendix), H takes the form 
\beq
H = \sum_{\k,\sigma} \epsilon_\k c_{\k,\sigma}^\dagger c_{\k,\sigma}
+ \sum_\q \omega_\q (\alpha_\q^\dagger \alpha_\q + 1/2) 
- I \int d\r \psi_\alpha^\dagger(\r) \sigma^z_{\alpha\beta} \psi_\beta (\r) 
\phi(\r)
\label{eqH}
\eeq

\noindent where 
\beq
\phi(\r)  =  \frac{1}{\sqrt{V}} \sum_\q 
\frac{1}{2} \lambda_\q \left( \alpha_\q + \alpha_{-\q}^\dagger \right)
e^{i\q\r};\qquad
\lambda_\q^2 = \frac{\Delta}{\omega_\q} .
\eeq

\noindent 
$\omega_\q$ is the dispersion of those bosons which have creation and
destruction operators $\alpha_\q^\dagger$ and $\alpha_\q$ respectively (see
the Appendix).  

We define the normal electron and exciton Green's functions as follows:
\beqa
G_{\alpha\beta}(\r-\r',\tau-\tau') & = & - \langle T \psi_\alpha(\r,\tau)
\overline{\psi}_\beta(\r',\tau') \rangle \\
D(\r-\r',\tau-\tau') & = & - \langle T \phi(\r,\tau)
\phi(\r',\tau') \rangle .
\eeqa

\noindent 
Note
that $\phi(\r) \sim S_x(\r)$, as can be seen immediately 
by comparing the terms $\sim I$ in Eqs.~(\ref{HfirstVers})
and~(\ref{eqH}), and so $D$ is essentially a pseudospin susceptibility. 
Furthermore, as in the theory of phonon-mediated superconductivity, 
$\phi(\r)$ is real, $\phi(\r) = \phi^\dagger(\r)$, and 
commutes with itself, $[\phi(\r),\phi(\r')] = 0$, conditions which allow us to
use Wick's Theorem. 
The \'{E}liashberg equations for the conduction-electron self-energy
$\Sigma(\p,i\omega_n)$ and gap function $\Phi(\p,i\omega_n)$ which follow
from the Hamiltonian Eq.~(\ref{eqH}) are 
\beqa
\Sigma(\p,i\omega_n) & = & \frac{T}{N} \sum_{\p'\omega_n'}
K(\p-\p',i\omega_n-i\omega_n') \, G(\p',i\omega_n') \\
\Phi(\p,i\omega_n) & = & p \, \frac{T}{N} \sum_{\p'\omega_n'}
K(\p-\p',i\omega_n-i\omega_n')
|G(\p',i\omega_n')|^2\, \Phi(\p',i\omega_n').
\label{gapEq}
\eeqa

\noindent 
$N$ is the total number of lattice sites. 
The kernel $K$ is given by 
\beq
K(\q,i\nu_n)  =  -I^2 D^0 ( \q, i\nu_n )
=\left( \frac{I^2\Delta}{2} \right) \frac{1}{\nu_n^2 + \omega_\q^2}.
\eeq

\noindent 
We have decided not to renormalise $D^0$ 
by the interaction $I$ between magnetic excitons 
and electrons. This interaction is, however, already included in $D^0$ to the
extent that we model the exciton dispersion $\omega_q$ 
by the true experimental dispersion.
A particularly noteworthy feature of this formulation is that 
the effective interaction is dominated by its static part and 
is strongly peaked in {\bf q}-space at the antiferromagnetic wavevector 
$\Q = (0,0,\frac{\pi}{c})$. This strong dependence on wavevector contrasts with
the practically wavevector-independent interaction in the usual
phonon-mediated superconductivity. The difference arises because, in the
phonon problem, the quantity
corresponding to $\lambda_\q$ has an additional factor,  
$\omega_\q$, leading to a wavevector-independent static phonon
propagator. 
A strong interaction  
between the collective modes of the localised moments and the heavy
conduction electrons  at $\Q = (0,0,\frac{\pi}{c})$ has actually been
observed\cite{Bernhoeft98,Bernhoeft00} in UPd$_2$Al$_3$. 
The electron Green's function is related to the electron self-energy via
the Dyson equation 
\beq
G^{-1}(\p,i\omega_n) = i\omega_n - (\epsilon_\p-\mu) - \Sigma(\p,i\omega_n) .
\eeq

The prefactor $p$ in Eq.~(\ref{gapEq}) 
is the expectation value of the Ising spin-spin interaction
$\hat{\sigma}^z\hat{\sigma}^z$
in the spin part of the pair wave function $|\chi\rangle = |S,\,S_z\rangle$
(${\bf S}$ here should not be confused with the pseudospin introduced earlier):
\beq
p = \langle\chi|  \hat{\sigma}^z\hat{\sigma}^z |\chi\rangle .
\eeq
In the opposite-spin pair (OSP) states, 
\beq
|\chi\rangle = 
\left\{ 
\begin{array}{l}
\frac{1}{\sqrt{2}} \left(
|\!\uparrow\downarrow\rangle - |\!\downarrow\uparrow\rangle \right)\\
\frac{1}{\sqrt{2}} \left( 
|\!\uparrow\downarrow\rangle + |\!\downarrow\uparrow\rangle \right)
\end{array} 
\right. ,
\eeq
p reduces to 
\beq
p = \langle\uparrow\downarrow\!| \hat{\sigma}^z\hat{\sigma}^z 
|\!\uparrow\downarrow\rangle 
= -1 .
\eeq
On the other hand, for each of the equal-spin pair (ESP) states,  
$ |\chi\rangle = 
|\!\!\uparrow\uparrow\rangle \mbox{ and }
|\!\!\downarrow\downarrow\rangle$,
p is $+1$. 
Note that spin rotational
symmetry is broken in a maximal (Ising-type) way in Eq.~(\ref{gapEq}). 
Consequently the usual classification of pairing states 
into an (even-parity) singlet with $S=0$ 
and an (odd-parity) triplet with $S=1$ is no longer valid.   
In our model, the $|1,0\rangle$ pair state 
no longer has the same energy as that of 
the $|1,\pm1\rangle$ pair states. 
The result is a different classification into 
an ESP doublet and two OSP singlet states belonging to irreducible
representations of the D$_{6h}$ group (Table~\ref{tab:gap}). 

Finally, the conduction-band filling, defined as the ratio of the number of
electrons to the number of lattice sites, is
\beq
n = \frac{1}{N} \sum_{\k} 
n_\k,
\eeq

\noindent 
where $n_\k$ is the quasiparticle occupation factor: 
\beqa
n_\k & = & n_{\k\uparrow} + n_{\k\downarrow} 
= 2 T \sum_{\omega_n} G ( \k, i\omega_n ) .
\eeqa

Experimentally, the magnetic exciton 
has its strongest dispersion $\omega_\q$ 
along the c-direction. Neglecting the weaker
dispersion in the a-b plane allows us to reduce the three-dimensional
problem to one dimension:

\beqa
\Sigma(p_z,i\omega_n) & = & \frac{T}{N_z} \sum_{p_z'\omega_n'}
K(p_z-p_z',i\omega_n-i\omega_n') \int \frac{d\p_\perp'}{\left(2\pi\right)^2}
G(\p_\perp',p_z',i\omega_n') \label{Eli1Da}\\
\Phi(p_z,i\omega_n) & = & p \, \frac{T}{N_z} \sum_{p_z'\omega_n'}
K(p_z-p_z',i\omega_n-i\omega_n')\, \Phi(p_z',i\omega_n') 
\int \frac{d\p_\perp'}{\left(2\pi\right)^2} 
|G(\p_\perp',p_z',i\omega_n')|^2 \label{Eli1Db}\\ 
n & = & \frac{1}{N_z} \sum_{k_z} A \int \frac{d\k_\perp}{\left(2\pi\right)^2}
n_\k, \label{Eli1Dc}
\eeqa

\noindent
provided we can carry out the integrals along the perpendicular
direction analytically. 
Here, $N_z$ is the number of lattice sites in the z-direction and $A$ is the
area of the (hexagonal) unit cell in the plane.  
We model the exciton dispersion by 
\beq
\begin{array}{cc}
\omega(q_z)  =\omega_{ex}[1 + \beta \cos ( q_z )]; & 0 < \beta\simeq 1  ,
\end{array}
\eeq

\noindent 
This form of the dispersion along the c-direction describes qualitatively the
observed excitation branch \cite{Mason97}. Here
2$\beta\omega_{ex}\simeq$ 8 meV is the overall dispersion width and
(1-$\beta)\omega_{ex}\simeq$ 1 meV is the exciton gap at the AF wave
vector {\bf Q}=(0,0,$\frac{\pi}{c}$). The quantity $\omega_{ex}$ is a
characteristic exciton energy. We choose $\beta$ = 0.8 and
$\omega_{ex}\simeq$ 5 meV $\simeq$ 60 K. The detailed RPA theory of
the magnetic-exciton dispersion and a fit to the INS data of
Ref.~\onlinecite{Mason97} was given in Ref.~\onlinecite{Thalmeier02}.

That sheet of the Fermi surface of UPd$_2$Al$_3$ thought to be
most important in bringing about heavy-fermion behaviour \cite{Thalmeier02}
has the shape of a corrugated cylinder with its axis aligned along the
c-direction of the hexagonal lattice. We therefore choose to model the
electron dispersion as the sum of a strongly dispersive part in the plane
$\epsilon_{\k_\perp}$ and a weakly dispersive part in the c-direction 
$\epsilon_{k_z}$.
We approximate the hexagonal
unit cell in the plane ($k_x, k_y$) by a circle with radius $k_0$, chosen 
so that the hexagon and circle have the same area.
(The areas must be the same in order
that the maximum value of the band filling is two in both cases.) 
Then, 
assuming a parabolic dispersion in the plane $\epsilon_{\k_\perp} =
\epsilon_\perp w^2$ ($0<w=k_\perp/k_0<1$),
we may carry out the integrals in
Eqs.~(\ref{Eli1Da}), (\ref{Eli1Db}) and (\ref{Eli1Dc})
analytically. 
In the following we measure all energies, temperatures and frequencies
in units of $\epsilon_\perp$. 
The reduced 
\'{E}liashberg equations are
\beqa
\Sigma(p_z,i\omega_n) & = & \frac{T}{N_z} \sum_{p_z'\omega_n'}
K(p_z-p_z',i\omega_n-i\omega_n') 
G_z(p_z',i\omega_n') \label{reducedSigma}\\
\lambda(T) \Phi(p_z,i\omega_n) & = & p \, \frac{T}{N_z} \sum_{p_z'\omega_n'}
K(p_z-p_z',i\omega_n-i\omega_n')\, M_z(p_z',i\omega_n') 
\Phi(p_z',i\omega_n') \label{reducedPhi} \\
n & = & 1 - \frac{2T}{N_z} \sum_{k_z,\omega_n>0} \ln 
\left[  \frac{(z'-1)^2 + (z'')^2}{(z')^2+(z'')^2} \right] ,
\eeqa

\noindent where 
\beqa
K(q_z,i\nu_n) & = & \frac{g}{
\left(\omega_{q_z}/\omega_{ex}\right)^2 
+
\left(\nu_n/\omega_{ex}\right)^2
} \\
G_z'(p_z',i\omega_n) & = & 
-\frac{1}{2} \ln \left[ \frac{(z'-1)^2 + (z'')^2}{(z')^2+(z'')^2}
\right] \\
G_z''(p_z',i\omega_n) & = &
- \left\{ \tan^{-1}\left( \frac{1-z'}{z''} \right) - 
\tan^{-1} \left( -\frac{z'}{z''} \right) \right\} \\
M_z(p_z',i\omega_n) & = & 
\frac{1}{z''} \left\{ \tan^{-1}\left( \frac{1-z'}{z''} \right) - 
\tan^{-1} \left( -\frac{z'}{z''} \right) \right\} .
\eeqa

\noindent
and 
\beq
\lambda(T) = 1 \mbox{ for } T = T_c.
\label{condTc}
\eeq

\noindent
We have lumped together 
numerical prefactors and the original coupling constant
$I$ into a new coupling constant $g$ having dimensions of energy: 

\beq
g  =  \frac{I^2\Delta}{2} \left( \frac{1}{2c} \frac{k_0^2}{2\pi} 
\right) \frac{1}{\omega_{ex}^2} ,
\label{coupling1}
\eeq

\noindent 
where $c$ is the lattice constant in the z-direction.
The value of $g$ is
not calculated; rather it is considered  
a model parameter motivated by experiment. 
The complex number $z$ has a real part
\beq
z' = -\left( \epsilon_{p_z'} - \mu + \Sigma'(p_z',i\omega_n) \right) 
\eeq

\noindent and an imaginary part
\beq
z'' = \omega_n - \Sigma''(p_z',i\omega_n) .
\eeq

\noindent 
We choose a simple tight-binding form for $\epsilon_{p_z'}$
\beq
\begin{array}{cc}
\epsilon_{p_z'} = \frac{\alpha}{2} \cos ( p_z' ); & \alpha \ll 1 
\end{array}
\eeq

\noindent
where $\alpha$ determines the degree of corrugation of the FS cylinder
along c.

The momentum convolutions in Eqs.~(\ref{reducedSigma}) and~(\ref{reducedPhi})
were evaluated with the aid of a fast-Fourier-transform algorithm on a 
$32\times 32$ lattice. The corresponding frequency sums were carried out 
using the renormalisation group technique of Pao and Bickers 
\cite{PaoBickers94} which allows a considerable reduction of the
computational effort.
Between $240$ and $480$ Matsubara frequencies were kept at each stage of the
renormalisation group procedure. The renormalisation procedure was started 
at a temperature $T_0=0.01 \epsilon_\perp$ 
and the frequency sum cut-off used was
$\Omega_c \approx 15 \epsilon_\perp$. 
The renormalisation procedure restricts us to 
discrete temperatures so that the point at which the condition in 
\mbox{Eq. (\ref{condTc})} is met must be determined by interpolation. 
The discrete temperatures were sufficiently close that a linear 
interpolation was adequate.

\section{Results and Discussion}
 
Scattering by isolated impurities with CEF 
excitations is usually pair breaking in the s-wave
channel\cite{FuldePeschel72} because the dipolar exchange interactions which
break the singlet state are stronger than spin-conserving quadrupolar
interactions between conduction electrons and CEF states.
The s-wave state (where $\Phi(p_z)$ is nodeless) 
is also not favourable in our case of a periodic
lattice of partly localised 5f-electrons which has dispersive CEF
excitations. One can most clearly see this in the following way. 
In the singlet channel $p=-1$. Because the kernel $K(\q,i\nu_n)$ is
strongly peaked at $q_z=\frac{\pi}{c}$ and $\nu_n = 0$, the gap
equation may then be crudely approximated in the following way:
\beq
\Phi(p_z,i\pi T) \simeq 
- \, C(p_z,\frac{\pi}{c}) \, \Phi(p_z-\frac{\pi}{c},i\pi T),
\eeq
where $C(p_z,\frac{\pi}{c})$ is a positive number. 
Therefore a finite gap $\Phi(p_z,i\pi T)$ must change sign on translation
through $Q_z = \frac{\pi}{c}$, i.e., the gap must possess the symmetry
 $\Phi(p_z-\frac{\pi}{c}) =
-\Phi(p_z)$, which excludes an s-wave gap function.

However a non-s-wave superconducting state with
nodes may take advantage of the exciton dispersion and become stable.
From numerical solutions of the linearised gap equations for our model
we find that the instability with the highest T$_c$ is accidentally 
doubly degenerate.
The corresponding gap functions transform as the even- and
odd-parity OSP states, $\cos(cp_z)$ and $\sin(cp_z)$. The former
is the usual singlet state; the latter is the S$_z$=0
part of the triplet which would appear in a theory with full spin rotational
symmetry. 
We also find an instability in the ESP channel but at a much lower
temperature. The corresponding pairing symmetry is $\Phi(p_z) \sim 
\sin(2cp_z)$. We summarise our results in Table~\ref{tab:gap}.

\begin{table}
\caption{Spin and orbital structure of the possible gap functions
which are solutions of the
\'{E}liashberg equations for the dual model of UPd$_2$Al$_3$ used in this 
paper.\label{tab:gap}
}
\begin{ruledtabular}
\begin{tabular}{lllll}
$p$ & $|\chi\rangle = |S,\,S_z\rangle$ & D$_{6h}$ repres. & spin
pairing & $\Phi(p_z)$ \\
\hline
-1 & $|0,0\rangle$ = $\frac{1}{\sqrt{2}} \left(
|\!\uparrow\downarrow\rangle -|\!\downarrow\uparrow\rangle \right)$ 
&$\Gamma_1^+$ (A$_{1g}$) & OSP & $\cos(cp_z)$ \\
-1 & $|1,0\rangle$ = $\frac{1}{\sqrt{2}} \left( |\!\uparrow\downarrow\rangle +
|\!\downarrow\uparrow\rangle \right)$
&$\Gamma_1^-$ (A$_{1u}$)  & OSP & $\sin(cp_z)$ \\
+1 & $|1,\pm1\rangle$ = $|\!\uparrow\uparrow\rangle,
|\!\downarrow\downarrow\rangle$ 
&$\Gamma_1^-$ (A$_{1u}$) & ESP & $\sin(2cp_z)$
\end{tabular}
\end{ruledtabular}
\end{table}

The symmetries of the orbital pair wavefunctions can be understood in the
following way. By comparing our self-energy equations with those obtained from
a four-fermion interaction we see that the effective interaction between
quasiparticles in our theory is 
\beq
\langle \gamma_2\gamma_4 | \hat{V} | \gamma_1\gamma_3 \rangle 
= I^2 D^0 \sigma^z_{\gamma_1\gamma_2} \sigma^z_{\gamma_3\gamma_4},
\eeq
or equivalently
\beq
\hat{V} = I^2 D^0 \hat{\sigma^z}\hat{\sigma^z}.
\eeq
Hence the pairing interaction in some $|\chi\rangle$ channel is
\beq
V_\chi(\q,i\nu_n) = \langle \chi | \hat{\sigma^z}\hat{\sigma^z} | \chi \rangle I^2 D^0(\q,i\nu_n) .
\eeq

\noindent 
We note that our equations go beyond previous
calculations \cite{Zwicknagl02,Thalmeier02} 
in that our formulation includes the full
momentum and frequency dependence of the effective interaction
$V_\chi(\q,i\nu_n)$. 

In the OSP channel the static interaction is
\beq
V(q_z)=-I^2 D^0(q_z) = \frac{I^2\Delta/2}{\omega(q_z)^2}, 
\eeq
which is strongly peaked at $\frac{\pi}{c}$. The corresponding interaction in real
space is therefore attractive when the quasiparticles are seperated by a
lattice spacing in the c-direction. Now $\cos(cp_z)$ and $\sin(cp_z)$
are peaked with equal amplitude in real space at $z=c$. Hence they 
are equally well suited to take advantage of the attractive
part of the interaction and have equal superconducting transition
temperatures. 
The ESP interaction is the negative of the OSP interaction and so it first
becomes attractive at
$z=2c$ in real space. 
The only odd-parity wavefunction which is peaked in real space 
at this position is
$\sin(2cp_z)$. 
 
Ultimately, 
the degeneracy of the OSP states is a peculiarity of the particular CEF level
scheme we adopt for the localised 5f states and of the approximate
form of their exciton dispersion which we assume (viz. strongly wavevector
dependent only in the z-direction). 
Both OSP states are, in fact, compatible with present experimental evidence. 
OSP states, regardless of the value of total spin angular momentum $S$, 
give rise to a reduction in the paramagnetic susceptibility on entering the
superconducting state. 
This reduction comes about because oppositely paired
spins have no magnetisation. Furthermore, such a reduction in the Pauli 
susceptibility leads to
an upper critical field $H_{c2}$ 
which, in some cases, can be smaller than that due to
the Meissner effect. This effect is called Pauli limiting \cite{Clogston62}.
A pronounced Knight-shift reduction at Pd
sites below $T_c$ (Refs.~\onlinecite{Tou03,Feyerherm94})
 and $H_{c2}$ Pauli-limiting 
\cite{Hessert97} have both been observed in UPd$_2$Al$_3$. Note also that
both OSP gap
functions have node lines perpendicular to the c-direction. 
In the even-parity case they
are located at $k_z = \pm \frac{1}{2} Q_z$, i.e., at the Bragg planes and
zone boundaries of the AF Brillouin zone, whereas in the odd-parity case
they are located at $k_z$=0. Experimentally, the existence of node lines was
inferred from NMR relaxation-rate \cite{Tou95,Matsuda97},
specific-heat \cite{Sato93} and thermal-conductivity measurements
\cite{Hiroi97}. These experiments did not, however, 
locate the node position along the c-direction
and so the correct gap symmetry 
in UPd$_2$Al$_3$ presently remains an open problem. To distinguish
between the possible nodal gap functions, it is very important to
perform field-angle dependent measurements of the specific heat
or of the thermal conductivity at low temperature. As
proposed in Ref.~\onlinecite{Thalmeier02a}, such measurements
may be able to
locate the node line in {\bf q}-space.

\begin{figure}
\includegraphics[width=12cm]{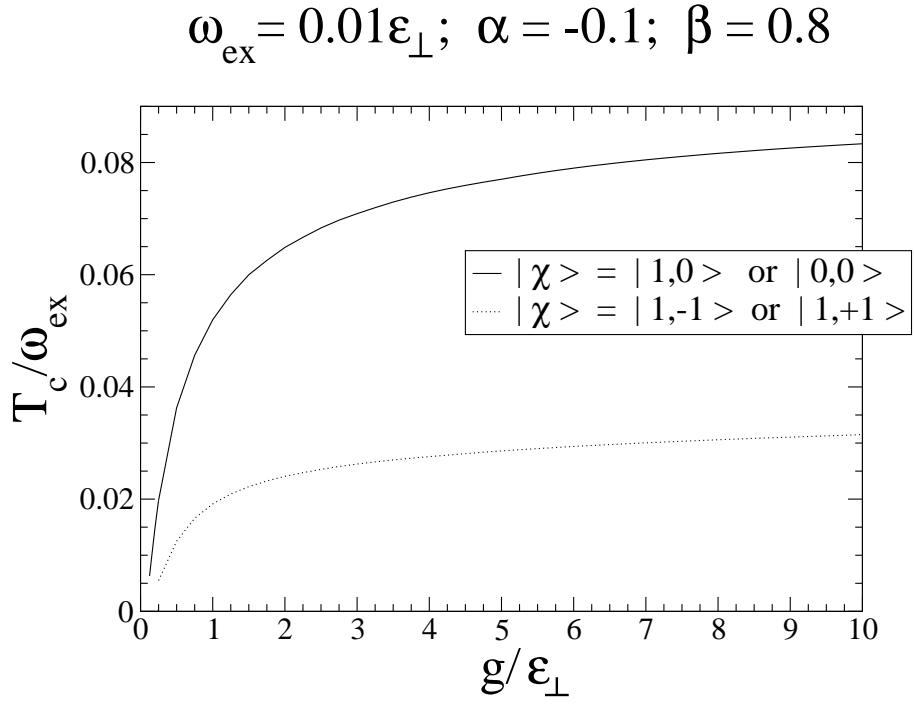}
\caption{\label{fig1}The dependence of the 
superconducting transition temperature $T_c$
on the coupling constant
$g$ [Eq.~(\ref{coupling1})]. 
The band filling $n$ is $0.6$.}
\end{figure}

The results of our numerical calculations of the superconducting transition
temperature in the ESP and OSP channels are shown in Fig.~\ref{fig1}. The
superconducting instability occurs first in the OSP channel. 
We have carried out
the calculations with $\omega_{ex} = 0.01\epsilon_{\perp}$. This exciton
energy corresponds to about $60$K for an electronic bandwidth of $0.5$ eV. 
One therefore obtains a transition temperature in the OSP channel close 
to the experimental value\cite{Geibel91} ($T_c = 1.8$K) 
with a dimensionless coupling constant $g/\epsilon_\perp$ of about $0.5$.
At this point we remind the reader that this temperature is in the range of 
validity of our theory
$T_c \ll \omega_{ex}  \sim \Delta$.

\begin{figure}
\includegraphics[width=12cm]{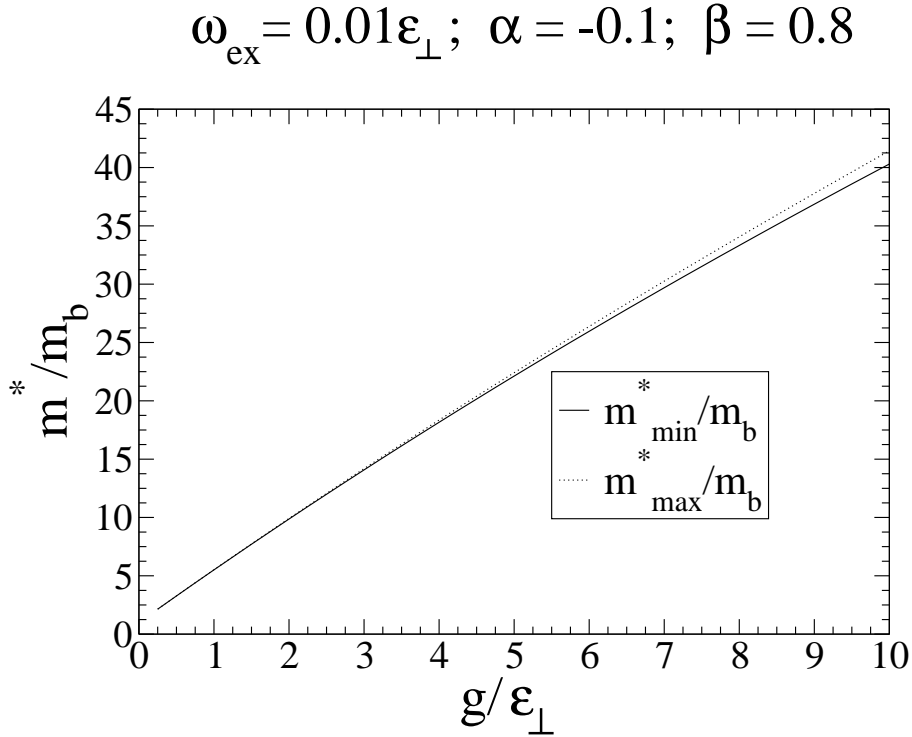}
\caption{\label{fig2}The dependence of the 
mass renormalisation $m^*/m_b$
on the coupling constant
$g$ [Eq.~(\ref{coupling1})].
The band filling $n$ is $0.6$ as in Fig.~\ref{fig1}. m$^*_{min}$
and m$^*_{max}$ are the extremal effective masses over all values of
q$_z$. m$^*_{min}$ occurs at q$_z$ = 0; m$^*_{max}$ occurs at
q$_z=\frac{\pi}{c}$. $m^*$ is practically constant over the Fermi
surface for the range of  values of $g$ which we consider in this work
($0.5 < g/\epsilon_\perp < 4$).} 
\end{figure}

That contribution to the
mass renormalisation $m^*/m_b$ ($m_b$ is the band mass)
arising from the momentum dependence of the real part of
the self energy is small and so $m^*/m_b$ simply reduces to the
\'{E}liashberg renormalisation
factor, $Z(\p,i\omega_n)$, which is practically momentum-independent. 
The weak dependence of $Z(\p,i\omega_n)$ on $\p$ 
follows from the combined effect of 
a strong exciton dispersion along the c-axis and a weak electron dispersion
in the same direction.
Our results for the mass renormalisation are shown in
Fig.~\ref{fig2}.
Note that the mass renormalisation is approximately linear in 
$g \sim I^2$ (see Ref.~\onlinecite{FuldeJensen83}).
\emph{Ab initio} calculations of the band masses $m_b/m_0$ ($m_0$ is the
free electron mass) for the 
$\gamma$ ring and $\beta$ ring of the cylindrical Fermi surface have been
carried out by confining two of the three 5f electrons to
the Uranium ions\cite{Zwicknagl03}. These values, in conjunction with a mass enhancement of 
$m^*/m_b$ of about $10$, yielded values of $m^*/m_0$ which were in good
agreement with experiment\cite{Inada99}. 
We find that such a large value of the mass
renormalisation corresponds to a value of the coupling constant 
($g/\epsilon_\perp\approx2$) which is 
roughly a factor of $4$ larger than that required to reproduce the
transition temperature ($g/\epsilon_\perp\approx 0.5$). One should note that 
the AF long-range order may act to reduce $T_c$, an effect which has not been
considered here. 
Therefore the proper $g$
may indeed be larger than $0.5$.

We have also carried out our calculations in the case that the Fermi surface
is purely cylindrical ($\alpha = 0$). The results are practically
unchanged. This case, however, has the appealing feature that the 
bracketed factor in
the definition of the coupling parameter (see Eq.~\ref{coupling1}) is related
in the following simple way to the (constant) density of states per spin,
$N(\omega)$,
\beq
g = \frac{I^2\Delta}{2} N(\mu)\epsilon_\perp \frac{1}{\omega_{ex}^2}.
\eeq

\noindent
The dimensionless coupling constant can now be written
\beq
g/\epsilon_\perp = \frac{1}{2} \left( \lambda_1\lambda_2 \right)^2 
\lambda_3 ,
\eeq

\noindent
where the dimensionless constants $\lambda_i$ are defined by 
\beqa
\lambda_1 & = & IN(\mu) \\
\lambda_2 & = & \Delta / \omega_{ex} \\
\lambda_3 & = & \frac{1}{\Delta N(\mu)} .
\eeqa

\noindent 
Of these three parameters, the first, $\lambda_1$, 
is not presently known for the
Uranium 5f systems. 
The second, $\lambda_2$, is about unity. The third,
$\lambda_3$, is approximately $167$, if we make the estimate $N(\mu) \approx
1$eV$^{-1}$ per cell per spin. The value of the coupling parameter we
require to get the experimentally observed transition temperature then leads
to the estimate $\lambda_1 \approx 0.08$. A value of $\lambda_1$ which is a 
factor of $2$ larger would yield a mass renormalisation in good agreement with
experiment, but then $T_c$ would be larger than the experimental value ($1.8
K$) 
by roughly a factor of 1.6. 
As already mentioned, this discrepancy may be alleviated by
the effects of AF order. The values for $\lambda_1$
obtained here for a dual 5f compound are of the same order
of magnitude as for 4f systems\cite{WhiteFulde81,FuldeJensen83}.

\section{Conclusions and Outlook}

Motivated by experiments on UPd$_2$Al$_3$, 
we have examined, for the first time, a magnetic-exciton model of
superconductivity within a strong-coupling retarded framework.
The model naturally explains the strong interaction  
between the collective modes of the localised moments and the heavy
conduction electrons which is observed 
at $\Q = (0,0,\frac{\pi}{c})$ in  UPd$_2$Al$_3$. 
Solutions of the \'{E}liashberg equations show that the model 
favours a superconducting
instability for the even- or odd-parity OSP states. Each state has line
nodes perpendicular to the c-direction and
total z-component of spin $S_z = 0$, characteristics which are compatible with 
measurements of UPd$_2$Al$_3$. We find that a 
superconducting transition temperature 
and mass renormalisation, each in good agreement with
their experimental values in UPd$_2$Al$_3$,  
can be obtained using reasonable values of the parameters in the theory.
Taken together, these results strengthen the argument in favour 
of a magnetic-exciton
system in UPd$_2$Al$_3$. 

The evidence for dual f-systems in other Uranium compounds 
does not weigh as heavily as it does in the case of UPd$_2$Al$_3$.
In UPt$_3$ the number of itinerant f-electrons remains controversial.
However band-structure calculations based on the assumption that two
of the three 5f-electrons are localised reproduce the observed de
Haas-van Alphen frequencies as well as the anisotropic heavy electron
masses in this compound very well\cite{Zwicknagl03}.  

Recently unconventional superconductivity with split-T$_c$
superconductivity has been
discovered in the Pr-based skutterudite PrOs$_4$Sb$_{12}$
\cite{Bauer02,Vollmer03}. Thermodynamic and transport measurements
suggest that the observed heavy-fermion state arises from the interaction of
electric quadrupole moments of the CEF-split 4f$^2$ states of Pr$^{3+}$  
with the conduction electrons\cite{Bauer02}. The 
ground state is probably a singlet and the first excited state 
is probably a triplet, 
approximately 0.5 meV higher, with large off-diagonal quadrupolar matrix
elements connecting the two states. The largeness of these matrix elements,
together with the very small singlet-triplet splitting, strongly suggests an
effective mass renormalisation via virtual CEF excitations.
Finally, experiments reveal that 
the superconducting order parameter is 
anisotropic, making this compound another candidate for
CEF-exciton-mediated superconductivity, but this time of quadrupolar nature.

UPd$_2$Al$_3$ provides a strong motivation for the study of dual
f-electron systems such as the one discussed in this work. We have considered
a dual model in which all parameters are fixed except the coupling constant. 
It will be interesting to investigate the dependence of the transition
temperature on conduction-band filling and on the form of the exciton
dispersion, and to look for a mechanism that can discriminate between
the even- and odd-parity superconducting OSP states.

\begin{acknowledgments}

P.M. would like to thank P. Monthoux for valuable discussions. P.T. would like
to thank K. Maki, Y. Matsuda, K. Izawa, T. Watanabe and G. Varelogiannis for
helpful discussions. The authors
acknowledge discussions with G. Zwicknagl and A. Yaresko.

\end{acknowledgments}

\appendix*

\section{}

In this appendix we sketch the derivation of Eq.~(\ref{eqH}).
Our starting
point is the set of Holstein-Primakoff transformations (valid for
$T \ll \Delta$),
\beqa
S_z & = & -\frac{1}{2} + a^\dagger a \\
\left.
\begin{array}{lll}
S^+ & = & a \\
S^- & = & a^\dagger
\end{array} 
\right\}
& \implies & 
\begin{array}{lll} 
S_x & = & \frac{1}{2} ( a + a^\dagger )
\end{array} 
\eeqa

\noindent 
Inserting these expressions into the second and third sums in 
Eq.~(\ref{HfirstVers}) and Fourier
transforming, we arrive at the following form,
\beq
\sum_\q \left[ W_1(\q)  
a_\q^\dagger a_\q - \frac{W_2(\q)}{2} (a_\q^\dagger a_{-\q}^\dagger 
+ a_\q a_{-\q} ) \right] 
- 
\frac{I}{2N} \sum_{\k\q} c_{\k\alpha}^\dagger \sigma^z_{\alpha\beta} 
c_{\k+\q\beta} \, ( a_\q + a_{-\q}^\dagger) , 
\label{HthirdVers}
\eeq

\noindent
where 
\beq
W_1(\q) = \Delta - W_2(\q).
\label{W1DeltaW2} 
\eeq

\noindent
The first sum 
 can be diagonalised\cite{Grover65}, 
while at the same time introducing only a factor
$\lambda_\q$ in the second sum. To do this we use the Bogoliubov transformation
\beq
\left(
\begin{array}{c}
\alpha_\k \\
\alpha_{-\k}^\dagger
\end{array} 
\right)
= 
\left(
\begin{array}{cc}
u_\k & v_\k \\
v_\k & u_\k 
\end{array} 
\right)
\left(
\begin{array}{c}
a_\k \\
a_{-\k}^\dagger
\end{array} 
\right),
\eeq

\noindent
where $u_\k = \cosh(\theta_\k)$ and $v_\k = \sinh(\theta_\k)$. If $\theta_\k$
is chosen such that 
\beq
-W_2(\k)/W_1(\k) = \tanh(2\theta_\k),
\label{EqGamma}
\eeq

\noindent
then (ignoring 
constant additive terms) 
the first sum in Eq.~(\ref{HthirdVers}) takes the form, 
\beq
\sum_\q \omega_\q ( \alpha_\q^\dagger \alpha_\q  + 1/2) 
\eeq

\noindent 
where the dispersion $\omega_\q$ satisfies
\beq
\omega_\q^2 = W_1^2(\q) - W_2^2(\q).
\label{EqOmega} 
\eeq

\noindent 
Finally, it is easy to show that
\beq
a_\q + a_{-\q}^\dagger = \lambda_\q ( \alpha_\q + \alpha_{-\q}^\dagger ),
\eeq

\noindent 
where 
\beq 
\lambda_\q = \cosh(\theta_\q) - \sinh(\theta_\q) .
\eeq

\noindent 
Using Eqs.~(\ref{W1DeltaW2}) and~(\ref{EqGamma}), together with the identity
\beq
\Delta - 2W_2(\q) = \frac{\omega_\q^2}{\Delta},
\eeq

\noindent
we can then show that 
\beq
\lambda_\q^2  =  \frac{\Delta}{\omega_q} .
\eeq

\noindent
The interaction term in Eq.~(\ref{eqH}) follows immediately. 

\bibliography{exciton1DPM5} 

\begin{thebibliography}{35}
\expandafter\ifx\csname natexlab\endcsname\relax\def\natexlab#1{#1}\fi
\expandafter\ifx\csname bibnamefont\endcsname\relax
  \def\bibnamefont#1{#1}\fi
\expandafter\ifx\csname bibfnamefont\endcsname\relax
  \def\bibfnamefont#1{#1}\fi
\expandafter\ifx\csname citenamefont\endcsname\relax
  \def\citenamefont#1{#1}\fi
\expandafter\ifx\csname url\endcsname\relax
  \def\url#1{\texttt{#1}}\fi
\expandafter\ifx\csname urlprefix\endcsname\relax\def\urlprefix{URL }\fi
\providecommand{\bibinfo}[2]{#2}
\providecommand{\eprint}[2][]{\url{#2}}

\bibitem[{\citenamefont{Hewson}(1993)}]{HewsonBook}
\bibinfo{author}{\bibfnamefont{A.~C.} \bibnamefont{Hewson}},
  \emph{\bibinfo{title}{The Kondo Problem to Heavy Fermions}}
  (\bibinfo{publisher}{Cambridge University Press}, \bibinfo{year}{1993}).

\bibitem[{\citenamefont{Fulde et~al.}(1988)\citenamefont{Fulde, Keller, and
  Zwicknagl}}]{Fulde88}
\bibinfo{author}{\bibfnamefont{P.}~\bibnamefont{Fulde}},
  \bibinfo{author}{\bibfnamefont{J.}~\bibnamefont{Keller}}, \bibnamefont{and}
  \bibinfo{author}{\bibfnamefont{G.}~\bibnamefont{Zwicknagl}}, in
  \emph{\bibinfo{booktitle}{Solid State Physics}} (\bibinfo{publisher}{Academic
  Press}, \bibinfo{address}{San Diego}, \bibinfo{year}{1988}),
  vol.~\bibinfo{volume}{41}, p.~\bibinfo{pages}{1}.

\bibitem[{\citenamefont{Mathur et~al.}(1998)\citenamefont{Mathur, Grosche,
  Julian, Walker, Freye, Haselwimmer, and Lonzarich}}]{Mathur98}
\bibinfo{author}{\bibfnamefont{N.~D.} \bibnamefont{Mathur}},
  \bibinfo{author}{\bibfnamefont{F.~M.} \bibnamefont{Grosche}},
  \bibinfo{author}{\bibfnamefont{S.~R.} \bibnamefont{Julian}},
  \bibinfo{author}{\bibfnamefont{I.~R.} \bibnamefont{Walker}},
  \bibinfo{author}{\bibfnamefont{D.~M.} \bibnamefont{Freye}},
  \bibinfo{author}{\bibfnamefont{R.~K.~W.} \bibnamefont{Haselwimmer}},
  \bibnamefont{and} \bibinfo{author}{\bibfnamefont{G.~G.}
  \bibnamefont{Lonzarich}}, \bibinfo{journal}{Nature}
  \textbf{\bibinfo{volume}{394}}, \bibinfo{pages}{39} (\bibinfo{year}{1998}).

\bibitem[{\citenamefont{Curro et~al.}(2003)\citenamefont{Curro, Sarrao,
  Thompson, Pagliuso, \v{S}. Kos, Abanov, and Pines}}]{Curro03}
\bibinfo{author}{\bibfnamefont{N.~J.} \bibnamefont{Curro}},
  \bibinfo{author}{\bibfnamefont{J.~L.} \bibnamefont{Sarrao}},
  \bibinfo{author}{\bibfnamefont{J.~D.} \bibnamefont{Thompson}},
  \bibinfo{author}{\bibfnamefont{P.~G.} \bibnamefont{Pagliuso}},
  \bibinfo{author}{\bibnamefont{\v{S}. Kos}},
  \bibinfo{author}{\bibfnamefont{A.}~\bibnamefont{Abanov}}, \bibnamefont{and}
  \bibinfo{author}{\bibfnamefont{D.}~\bibnamefont{Pines}},
  \bibinfo{journal}{Phys. Rev. Lett.} \textbf{\bibinfo{volume}{90}},
  \bibinfo{pages}{227202} (\bibinfo{year}{2003}).

\bibitem[{\citenamefont{Petrovic et~al.}(2001)\citenamefont{Petrovic, Pagliuso,
  Hundley, Movshovich, Sarrao, Thompson, Fisk, and Monthoux}}]{Petrovic01}
\bibinfo{author}{\bibfnamefont{C.}~\bibnamefont{Petrovic}},
  \bibinfo{author}{\bibfnamefont{P.~G.} \bibnamefont{Pagliuso}},
  \bibinfo{author}{\bibfnamefont{M.~F.} \bibnamefont{Hundley}},
  \bibinfo{author}{\bibfnamefont{R.}~\bibnamefont{Movshovich}},
  \bibinfo{author}{\bibfnamefont{J.~L.} \bibnamefont{Sarrao}},
  \bibinfo{author}{\bibfnamefont{J.~D.} \bibnamefont{Thompson}},
  \bibinfo{author}{\bibfnamefont{Z.}~\bibnamefont{Fisk}}, \bibnamefont{and}
  \bibinfo{author}{\bibfnamefont{P.}~\bibnamefont{Monthoux}},
  \bibinfo{journal}{J. Phys. Condens. Matter} \textbf{\bibinfo{volume}{13}},
  \bibinfo{pages}{L337} (\bibinfo{year}{2001}).

\bibitem[{\citenamefont{Zwicknagl et~al.}(2002)\citenamefont{Zwicknagl,
  Yaresko, and Fulde}}]{Zwicknagl02}
\bibinfo{author}{\bibfnamefont{G.}~\bibnamefont{Zwicknagl}},
  \bibinfo{author}{\bibfnamefont{A.~N.} \bibnamefont{Yaresko}},
  \bibnamefont{and} \bibinfo{author}{\bibfnamefont{P.}~\bibnamefont{Fulde}},
  \bibinfo{journal}{Phys. Rev. B} \textbf{\bibinfo{volume}{65}},
  \bibinfo{pages}{081103} (\bibinfo{year}{2002}).

\bibitem[{\citenamefont{Zwicknagl et~al.}(2003)\citenamefont{Zwicknagl,
  Yaresko, and Fulde}}]{Zwicknagl03}
\bibinfo{author}{\bibfnamefont{G.}~\bibnamefont{Zwicknagl}},
  \bibinfo{author}{\bibfnamefont{A.}~\bibnamefont{Yaresko}}, \bibnamefont{and}
  \bibinfo{author}{\bibfnamefont{P.}~\bibnamefont{Fulde}},
  \bibinfo{journal}{Phys. Rev. B} \textbf{\bibinfo{volume}{68}},
  \bibinfo{pages}{052508} (\bibinfo{year}{2003}).

\bibitem[{\citenamefont{Fulde and Peschel}(1972)}]{FuldePeschel72}
\bibinfo{author}{\bibfnamefont{P.}~\bibnamefont{Fulde}} \bibnamefont{and}
  \bibinfo{author}{\bibfnamefont{I.}~\bibnamefont{Peschel}},
  \bibinfo{journal}{Adv. Phys.} \textbf{\bibinfo{volume}{21}},
  \bibinfo{pages}{1} (\bibinfo{year}{1972}).

\bibitem[{\citenamefont{White and Fulde}(1981)}]{WhiteFulde81}
\bibinfo{author}{\bibfnamefont{R.~M.} \bibnamefont{White}} \bibnamefont{and}
  \bibinfo{author}{\bibfnamefont{P.}~\bibnamefont{Fulde}},
  \bibinfo{journal}{Phys. Rev. Lett.} \textbf{\bibinfo{volume}{47}},
  \bibinfo{pages}{1540} (\bibinfo{year}{1981}).

\bibitem[{\citenamefont{Fulde et~al.}(1970)\citenamefont{Fulde, Hirst, and
  Luther}}]{Fulde70}
\bibinfo{author}{\bibfnamefont{P.}~\bibnamefont{Fulde}},
  \bibinfo{author}{\bibfnamefont{L.~L.} \bibnamefont{Hirst}}, \bibnamefont{and}
  \bibinfo{author}{\bibfnamefont{A.}~\bibnamefont{Luther}},
  \bibinfo{journal}{Z. Phys.} \textbf{\bibinfo{volume}{230}},
  \bibinfo{pages}{155} (\bibinfo{year}{1970}).

\bibitem[{\citenamefont{Geibel et~al.}(1991)\citenamefont{Geibel, Schank,
  Thies, Kitazawa, Bredl, B{\"{o}}hm, Rau, Grauel, Caspary, Helfrich
  et~al.}}]{Geibel91}
\bibinfo{author}{\bibfnamefont{C.}~\bibnamefont{Geibel}},
  \bibinfo{author}{\bibfnamefont{C.}~\bibnamefont{Schank}},
  \bibinfo{author}{\bibfnamefont{S.}~\bibnamefont{Thies}},
  \bibinfo{author}{\bibfnamefont{H.}~\bibnamefont{Kitazawa}},
  \bibinfo{author}{\bibfnamefont{C.~D.} \bibnamefont{Bredl}},
  \bibinfo{author}{\bibfnamefont{A.}~\bibnamefont{B{\"{o}}hm}},
  \bibinfo{author}{\bibfnamefont{M.}~\bibnamefont{Rau}},
  \bibinfo{author}{\bibfnamefont{A.}~\bibnamefont{Grauel}},
  \bibinfo{author}{\bibfnamefont{R.}~\bibnamefont{Caspary}},
  \bibinfo{author}{\bibfnamefont{R.}~\bibnamefont{Helfrich}},
  \bibnamefont{et~al.}, \bibinfo{journal}{Z. Phys. B}
  \textbf{\bibinfo{volume}{84}}, \bibinfo{pages}{1} (\bibinfo{year}{1991}).

\bibitem[{\citenamefont{Grauel et~al.}(1992)\citenamefont{Grauel, B{\"{o}}hm,
  Fischer, Geibel, K{\"{o}}hler, Modler, Schank, Steglich, and
  Weber}}]{Grauel92}
\bibinfo{author}{\bibfnamefont{A.}~\bibnamefont{Grauel}},
  \bibinfo{author}{\bibfnamefont{A.}~\bibnamefont{B{\"{o}}hm}},
  \bibinfo{author}{\bibfnamefont{H.}~\bibnamefont{Fischer}},
  \bibinfo{author}{\bibfnamefont{C.}~\bibnamefont{Geibel}},
  \bibinfo{author}{\bibfnamefont{R.}~\bibnamefont{K{\"{o}}hler}},
  \bibinfo{author}{\bibfnamefont{R.}~\bibnamefont{Modler}},
  \bibinfo{author}{\bibfnamefont{C.}~\bibnamefont{Schank}},
  \bibinfo{author}{\bibfnamefont{F.}~\bibnamefont{Steglich}}, \bibnamefont{and}
  \bibinfo{author}{\bibfnamefont{G.}~\bibnamefont{Weber}},
  \bibinfo{journal}{Phys. Rev. B} \textbf{\bibinfo{volume}{46}},
  \bibinfo{pages}{5818} (\bibinfo{year}{1992}).

\bibitem[{\citenamefont{Feyerherm et~al.}(1994)\citenamefont{Feyerherm, Amato,
  Gygax, Schenck, Geibel, Steglich, Sato, and Komatsubara}}]{Feyerherm94}
\bibinfo{author}{\bibfnamefont{R.}~\bibnamefont{Feyerherm}},
  \bibinfo{author}{\bibfnamefont{A.}~\bibnamefont{Amato}},
  \bibinfo{author}{\bibfnamefont{F.~N.} \bibnamefont{Gygax}},
  \bibinfo{author}{\bibfnamefont{A.}~\bibnamefont{Schenck}},
  \bibinfo{author}{\bibfnamefont{C.}~\bibnamefont{Geibel}},
  \bibinfo{author}{\bibfnamefont{F.}~\bibnamefont{Steglich}},
  \bibinfo{author}{\bibfnamefont{N.}~\bibnamefont{Sato}}, \bibnamefont{and}
  \bibinfo{author}{\bibfnamefont{T.}~\bibnamefont{Komatsubara}},
  \bibinfo{journal}{Phys. Rev. Lett.} \textbf{\bibinfo{volume}{73}},
  \bibinfo{pages}{1849} (\bibinfo{year}{1994}).

\bibitem[{\citenamefont{Dressel et~al.}(2002)\citenamefont{Dressel, Kasper,
  Pethukhov, Gorshunov, Gr{\"{u}}ner, Huth, and Adrian}}]{Dressel02}
\bibinfo{author}{\bibfnamefont{M.}~\bibnamefont{Dressel}},
  \bibinfo{author}{\bibfnamefont{N.}~\bibnamefont{Kasper}},
  \bibinfo{author}{\bibfnamefont{K.}~\bibnamefont{Pethukhov}},
  \bibinfo{author}{\bibfnamefont{B.}~\bibnamefont{Gorshunov}},
  \bibinfo{author}{\bibfnamefont{G.}~\bibnamefont{Gr{\"{u}}ner}},
  \bibinfo{author}{\bibfnamefont{M.}~\bibnamefont{Huth}}, \bibnamefont{and}
  \bibinfo{author}{\bibfnamefont{H.}~\bibnamefont{Adrian}},
  \bibinfo{journal}{Phys. Rev. Lett.} \textbf{\bibinfo{volume}{88}},
  \bibinfo{pages}{186404} (\bibinfo{year}{2002}).

\bibitem[{\citenamefont{Mason and Aeppli}(1997)}]{Mason97}
\bibinfo{author}{\bibfnamefont{T.~E.} \bibnamefont{Mason}} \bibnamefont{and}
  \bibinfo{author}{\bibfnamefont{G.}~\bibnamefont{Aeppli}},
  \bibinfo{journal}{Matematisk-fysiske Meddelelser}
  \textbf{\bibinfo{volume}{45}}, \bibinfo{pages}{231} (\bibinfo{year}{1997}).

\bibitem[{\citenamefont{Sato et~al.}(1997)\citenamefont{Sato, Aso, Lander,
  Roessli, Komatsubara, and Endoh}}]{Sato97}
\bibinfo{author}{\bibfnamefont{N.}~\bibnamefont{Sato}},
  \bibinfo{author}{\bibfnamefont{N.}~\bibnamefont{Aso}},
  \bibinfo{author}{\bibfnamefont{G.~H.} \bibnamefont{Lander}},
  \bibinfo{author}{\bibfnamefont{B.}~\bibnamefont{Roessli}},
  \bibinfo{author}{\bibfnamefont{T.}~\bibnamefont{Komatsubara}},
  \bibnamefont{and} \bibinfo{author}{\bibfnamefont{Y.}~\bibnamefont{Endoh}},
  \bibinfo{journal}{J. Phys. Soc. Japan} \textbf{\bibinfo{volume}{66}},
  \bibinfo{pages}{1884} (\bibinfo{year}{1997}).

\bibitem[{\citenamefont{Bernhoeft et~al.}(1998)\citenamefont{Bernhoeft, Sato,
  Roessli, Aso, Hiess, Lander, Endoh, and Komatsubara}}]{Bernhoeft98}
\bibinfo{author}{\bibfnamefont{N.}~\bibnamefont{Bernhoeft}},
  \bibinfo{author}{\bibfnamefont{N.}~\bibnamefont{Sato}},
  \bibinfo{author}{\bibfnamefont{B.}~\bibnamefont{Roessli}},
  \bibinfo{author}{\bibfnamefont{N.}~\bibnamefont{Aso}},
  \bibinfo{author}{\bibfnamefont{A.}~\bibnamefont{Hiess}},
  \bibinfo{author}{\bibfnamefont{G.~H.} \bibnamefont{Lander}},
  \bibinfo{author}{\bibfnamefont{Y.}~\bibnamefont{Endoh}}, \bibnamefont{and}
  \bibinfo{author}{\bibfnamefont{T.}~\bibnamefont{Komatsubara}},
  \bibinfo{journal}{Phys. Rev. Lett.} \textbf{\bibinfo{volume}{81}},
  \bibinfo{pages}{4244} (\bibinfo{year}{1998}).

\bibitem[{\citenamefont{Bernhoeft}(2000)}]{Bernhoeft00}
\bibinfo{author}{\bibfnamefont{N.}~\bibnamefont{Bernhoeft}},
  \bibinfo{journal}{Eur. Phys. J. B} \textbf{\bibinfo{volume}{13}},
  \bibinfo{pages}{685} (\bibinfo{year}{2000}).

\bibitem[{\citenamefont{Jourdan et~al.}(1999)\citenamefont{Jourdan, Huth, and
  Adrian}}]{Jourdan99}
\bibinfo{author}{\bibfnamefont{M.}~\bibnamefont{Jourdan}},
  \bibinfo{author}{\bibfnamefont{M.}~\bibnamefont{Huth}}, \bibnamefont{and}
  \bibinfo{author}{\bibfnamefont{H.}~\bibnamefont{Adrian}},
  \bibinfo{journal}{Nature} \textbf{\bibinfo{volume}{398}}, \bibinfo{pages}{47}
  (\bibinfo{year}{1999}).

\bibitem[{\citenamefont{Sato et~al.}(2001)\citenamefont{Sato, Aso, Miyake,
  Shiina, Thalmeier, Varelogiannis, Geibel, Steglich, Fulde, and
  Komatsubara}}]{Sato01}
\bibinfo{author}{\bibfnamefont{N.~K.} \bibnamefont{Sato}},
  \bibinfo{author}{\bibfnamefont{N.}~\bibnamefont{Aso}},
  \bibinfo{author}{\bibfnamefont{K.}~\bibnamefont{Miyake}},
  \bibinfo{author}{\bibfnamefont{R.}~\bibnamefont{Shiina}},
  \bibinfo{author}{\bibfnamefont{P.}~\bibnamefont{Thalmeier}},
  \bibinfo{author}{\bibfnamefont{G.}~\bibnamefont{Varelogiannis}},
  \bibinfo{author}{\bibfnamefont{C.}~\bibnamefont{Geibel}},
  \bibinfo{author}{\bibfnamefont{F.}~\bibnamefont{Steglich}},
  \bibinfo{author}{\bibfnamefont{P.}~\bibnamefont{Fulde}}, \bibnamefont{and}
  \bibinfo{author}{\bibfnamefont{T.}~\bibnamefont{Komatsubara}},
  \bibinfo{journal}{Nature} \textbf{\bibinfo{volume}{410}},
  \bibinfo{pages}{340} (\bibinfo{year}{2001}).

\bibitem[{\citenamefont{Thalmeier}(2002)}]{Thalmeier02}
\bibinfo{author}{\bibfnamefont{P.}~\bibnamefont{Thalmeier}},
  \bibinfo{journal}{Eur. Phys. J. B} \textbf{\bibinfo{volume}{27}},
  \bibinfo{pages}{29} (\bibinfo{year}{2002}).

\bibitem[{\citenamefont{Pao and Bickers}(1994)}]{PaoBickers94}
\bibinfo{author}{\bibfnamefont{C.-H.} \bibnamefont{Pao}} \bibnamefont{and}
  \bibinfo{author}{\bibfnamefont{N.~E.} \bibnamefont{Bickers}},
  \bibinfo{journal}{Phys. Rev. B} \textbf{\bibinfo{volume}{49}},
  \bibinfo{pages}{1586} (\bibinfo{year}{1994}).

\bibitem[{\citenamefont{Clogston}(1962)}]{Clogston62}
\bibinfo{author}{\bibfnamefont{A.~M.} \bibnamefont{Clogston}},
  \bibinfo{journal}{Phys. Rev. Lett.} \textbf{\bibinfo{volume}{9}},
  \bibinfo{pages}{266} (\bibinfo{year}{1962}).

\bibitem[{\citenamefont{Tou et~al.}(2003)\citenamefont{Tou, Ishida, and
  Kitaoka}}]{Tou03}
\bibinfo{author}{\bibfnamefont{H.}~\bibnamefont{Tou}},
  \bibinfo{author}{\bibfnamefont{K.}~\bibnamefont{Ishida}}, \bibnamefont{and}
  \bibinfo{author}{\bibfnamefont{Y.}~\bibnamefont{Kitaoka}},
  \bibinfo{journal}{cond-mat/0308562}  (\bibinfo{year}{2003}).

\bibitem[{\citenamefont{Hessert et~al.}(1997)\citenamefont{Hessert, Huth,
  Jourdan, Adrian, Rieck, and Scharnberg}}]{Hessert97}
\bibinfo{author}{\bibfnamefont{J.}~\bibnamefont{Hessert}},
  \bibinfo{author}{\bibfnamefont{M.}~\bibnamefont{Huth}},
  \bibinfo{author}{\bibfnamefont{M.}~\bibnamefont{Jourdan}},
  \bibinfo{author}{\bibfnamefont{H.}~\bibnamefont{Adrian}},
  \bibinfo{author}{\bibfnamefont{C.~T.} \bibnamefont{Rieck}}, \bibnamefont{and}
  \bibinfo{author}{\bibfnamefont{K.}~\bibnamefont{Scharnberg}},
  \bibinfo{journal}{Physica B} \textbf{\bibinfo{volume}{230-232}},
  \bibinfo{pages}{373} (\bibinfo{year}{1997}).

\bibitem[{\citenamefont{Tou et~al.}(1995)\citenamefont{Tou, Kitaoka, Asayama,
  Geibel, Schank, and Steglich}}]{Tou95}
\bibinfo{author}{\bibfnamefont{H.}~\bibnamefont{Tou}},
  \bibinfo{author}{\bibfnamefont{Y.}~\bibnamefont{Kitaoka}},
  \bibinfo{author}{\bibfnamefont{K.}~\bibnamefont{Asayama}},
  \bibinfo{author}{\bibfnamefont{C.}~\bibnamefont{Geibel}},
  \bibinfo{author}{\bibfnamefont{C.}~\bibnamefont{Schank}}, \bibnamefont{and}
  \bibinfo{author}{\bibfnamefont{F.}~\bibnamefont{Steglich}},
  \bibinfo{journal}{Phys. Soc. Jpn.} \textbf{\bibinfo{volume}{64}},
  \bibinfo{pages}{725} (\bibinfo{year}{1995}).

\bibitem[{\citenamefont{Matsuda et~al.}(1997)\citenamefont{Matsuda, Kohori, and
  Kohara}}]{Matsuda97}
\bibinfo{author}{\bibfnamefont{K.}~\bibnamefont{Matsuda}},
  \bibinfo{author}{\bibfnamefont{Y.}~\bibnamefont{Kohori}}, \bibnamefont{and}
  \bibinfo{author}{\bibfnamefont{T.}~\bibnamefont{Kohara}},
  \bibinfo{journal}{Phys. Rev. B} \textbf{\bibinfo{volume}{55}},
  \bibinfo{pages}{15223} (\bibinfo{year}{1997}).

\bibitem[{\citenamefont{Sato}()}]{Sato93}
\bibinfo{author}{\bibfnamefont{N.}~\bibnamefont{Sato}},
  \bibinfo{note}{unpublished}.

\bibitem[{\citenamefont{Hiroi et~al.}(1997)\citenamefont{Hiroi, Sera,
  Kobayashi, Haga, Yamamoto, and Onuki}}]{Hiroi97}
\bibinfo{author}{\bibfnamefont{M.}~\bibnamefont{Hiroi}},
  \bibinfo{author}{\bibfnamefont{M.}~\bibnamefont{Sera}},
  \bibinfo{author}{\bibfnamefont{N.}~\bibnamefont{Kobayashi}},
  \bibinfo{author}{\bibfnamefont{Y.}~\bibnamefont{Haga}},
  \bibinfo{author}{\bibfnamefont{E.}~\bibnamefont{Yamamoto}}, \bibnamefont{and}
  \bibinfo{author}{\bibfnamefont{Y.}~\bibnamefont{Onuki}}, \bibinfo{journal}{J.
  Phys. Soc. Jpn.} \textbf{\bibinfo{volume}{66}}, \bibinfo{pages}{1595}
  (\bibinfo{year}{1997}).

\bibitem[{\citenamefont{Thalmeier and Maki}(2002)}]{Thalmeier02a}
\bibinfo{author}{\bibfnamefont{P.}~\bibnamefont{Thalmeier}} \bibnamefont{and}
  \bibinfo{author}{\bibfnamefont{K.}~\bibnamefont{Maki}},
  \bibinfo{journal}{Europhys. Lett.} \textbf{\bibinfo{volume}{58}},
  \bibinfo{pages}{119} (\bibinfo{year}{2002}).

\bibitem[{\citenamefont{Jensen}(1983)}]{FuldeJensen83}
\bibinfo{author}{\bibfnamefont{P.~F.~J.} \bibnamefont{Jensen}},
  \bibinfo{journal}{Phys. Rev. B} \textbf{\bibinfo{volume}{27}},
  \bibinfo{pages}{4085} (\bibinfo{year}{1983}).

\bibitem[{\citenamefont{Inada et~al.}(1999)\citenamefont{Inada, Yamagami, Haga,
  Sakurai, Tokiwa, Honma, Yamamoto, \={O}nuki, and Yanagisawa}}]{Inada99}
\bibinfo{author}{\bibfnamefont{Y.}~\bibnamefont{Inada}},
  \bibinfo{author}{\bibfnamefont{H.}~\bibnamefont{Yamagami}},
  \bibinfo{author}{\bibfnamefont{Y.}~\bibnamefont{Haga}},
  \bibinfo{author}{\bibfnamefont{K.}~\bibnamefont{Sakurai}},
  \bibinfo{author}{\bibfnamefont{Y.}~\bibnamefont{Tokiwa}},
  \bibinfo{author}{\bibfnamefont{T.}~\bibnamefont{Honma}},
  \bibinfo{author}{\bibfnamefont{E.}~\bibnamefont{Yamamoto}},
  \bibinfo{author}{\bibfnamefont{Y.}~\bibnamefont{\={O}nuki}},
  \bibnamefont{and}
  \bibinfo{author}{\bibfnamefont{T.}~\bibnamefont{Yanagisawa}},
  \bibinfo{journal}{J. Phys. Soc. Japan} \textbf{\bibinfo{volume}{68}},
  \bibinfo{pages}{3643} (\bibinfo{year}{1999}).

\bibitem[{\citenamefont{Bauer et~al.}(2002)\citenamefont{Bauer, Frederick, Ho,
  Zapf, and Maple}}]{Bauer02}
\bibinfo{author}{\bibfnamefont{E.~D.} \bibnamefont{Bauer}},
  \bibinfo{author}{\bibfnamefont{N.~A.} \bibnamefont{Frederick}},
  \bibinfo{author}{\bibfnamefont{P.-C.} \bibnamefont{Ho}},
  \bibinfo{author}{\bibfnamefont{V.~S.} \bibnamefont{Zapf}}, \bibnamefont{and}
  \bibinfo{author}{\bibfnamefont{M.~B.} \bibnamefont{Maple}},
  \bibinfo{journal}{Phys. Rev. B} \textbf{\bibinfo{volume}{65}},
  \bibinfo{pages}{100506} (\bibinfo{year}{2002}).

\bibitem[{\citenamefont{Vollmer et~al.}(2003)\citenamefont{Vollmer, Faisst,
  Pfleiderer, v.~L\"ohneysen, Bauer, Ho, and Maple}}]{Vollmer03}
\bibinfo{author}{\bibfnamefont{R.}~\bibnamefont{Vollmer}},
  \bibinfo{author}{\bibfnamefont{A.}~\bibnamefont{Faisst}},
  \bibinfo{author}{\bibfnamefont{C.}~\bibnamefont{Pfleiderer}},
  \bibinfo{author}{\bibfnamefont{H.}~\bibnamefont{v.~L\"ohneysen}},
  \bibinfo{author}{\bibfnamefont{E.~D.} \bibnamefont{Bauer}},
  \bibinfo{author}{\bibfnamefont{P.~C.} \bibnamefont{Ho}}, \bibnamefont{and}
  \bibinfo{author}{\bibfnamefont{M.~B.} \bibnamefont{Maple}},
  \bibinfo{journal}{Phys. Rev. Lett.} \textbf{\bibinfo{volume}{90}},
  \bibinfo{pages}{57001} (\bibinfo{year}{2003}).

\bibitem[{\citenamefont{Grover}(1965)}]{Grover65}
\bibinfo{author}{\bibfnamefont{B.}~\bibnamefont{Grover}},
  \bibinfo{journal}{Phys. Rev.} \textbf{\bibinfo{volume}{140}},
  \bibinfo{pages}{A1944} (\bibinfo{year}{1965}).

\end{thebibliography}

\end{document}